\documentclass[prl,aps, babel,twocolumn]{revtex4}
\usepackage{amsmath,amssymb,amstext,amscd,amsthm,amsopn,graphicx,indentfirst,mathrsfs}

\graphicspath{{.}{./EPS/}}

\begin{document}

\title{Quantum zigzag transition in ion chains}

\author{Efrat Shimshoni,$^1$ Giovanna Morigi,$^{2,3}$ and Shmuel Fishman$^4$}

\affiliation{
$^1$ Department of Physics, Bar-Ilan University, Ramat-Gan 52900, Israel\\
$^2$ Theoretische Physik, Universit{\"a}t des Saarlandes, D 66041 Saarbr{\"u}cken, Germany\\
$^3$ Department de F\'isica, Universitat Aut\`onoma de Barcelona, E 08193 Bellaterra, Spain\\
$^4$ Department of Physics, Technion, Haifa 32000, Israel
}

\date{\today}

\begin{abstract}

A string of trapped ions at zero temperature exhibits a structural
phase transition to a zigzag structure, tuned by reducing the
transverse trap potential or the interparticle distance. The
transition is driven by transverse, short wavelength vibrational
modes. We argue that
 this  is a quantum phase transition, which can be experimentally realized and probed.
 Indeed, by means of a mapping to the Ising model in a transverse field, we estimate
the quantum critical point in terms of the system parameters, and find a finite, measurable deviation from the critical point
predicted by the classical theory. A measurement procedure is suggested which can probe the effects of quantum fluctuations at
criticality. These results can be extended to describe the transverse instability of ultracold polar molecules in a one
dimensional optical lattice.
\end{abstract}


\maketitle

Crystals of singly-charged ions, ion Coulomb crystals, constitute
a fascinating physical realization of the phenomenon discussed by
Wigner for electrons in metals~\cite{Wigner}. These crystals
result from the combined Coulomb repulsive force and the restoring
force of an external potential, usually a harmonic
trap~\cite{Dubin}. The strongly-correlated regime, leading to
long-range order, is accessed by means of laser cooling, allowing
one to obtain samples at temperatures
$T$
ranging from few mK
to hundreds of $\mu$K~\cite{Eschner}.


Ion chains have been most discussed in the literature focussed on quantum technological applications~\cite{Cirac,Ibk_Review}. They are realized in potentials with an enhanced elliptical geometry, so that the charges align along the trap axis~\cite{Walther,Wineland,Drewsen}. Strictly speaking, in one dimension there is no crystalline order even at ultralow temperatures. However the Coulomb interaction gives rise to very-slowly decaying density-density correlations, such that quasi long-range order is generally warranted~\cite{Schulz93}. At a given interparticle distance the stability of the ion chain is determined by the strength of the transverse potential: if the corresponding frequency $\nu_t$ is below a critical value $\nu_c$, a planar, zigzag structure is observed~\cite{Hasse,Walther,Wineland,Dubin93}. In Ref.~\cite{Piacente, FCCM} it was shown that the classical structural transition linear-zigzag is of second-order.
A Landau model was derived, proving that the soft mode of the phase transition is the zigzag mode of the linear chain (the transverse mode with the shortest wavelength) which drives the instability and determines the new structure~\cite{FCCM}. The corresponding Ginzburg-Landau equation in the continuum limit is reported in~\cite{DelCampo}. The results of these studies are strictly valid in the classical regime, when quantum and thermal fluctuations can be neglected.

\begin{figure}
\includegraphics[angle=0, width=0.8 \linewidth, height=!]{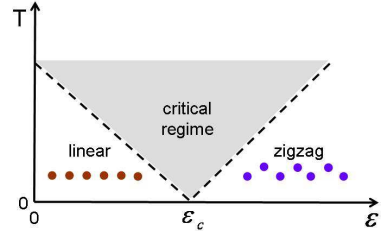}
\caption{
Phase diagram for a linear-zigzag transition, according to the
mapping to the quantum Ising transition in 1D, where $T$ is the
sample temperature and the dimensionless parameter $\varepsilon$
is tuned by the confining potential or the interparticle distance.
The classical instability of the linear chain~\cite{FCCM} occurs
at $\varepsilon=0$. The quantum critical point, at
$\varepsilon_c>0$, separates the linear from the zigzag phase at
$T=0$. For $0<\varepsilon<\varepsilon_c$ quantum fluctuations
dominate, and the crystal is in the linear (disordered) phase. The
dashed lines indicate the boundaries of the quantum critical
region. }\label{Fig:1}
\end{figure}

A question naturally emerges whether and how quantum fluctuations
modify the behavior at the linear-zigzag instability. Theoretical
studies on ion systems so far focused on
small samples~\cite{Retzker}. In this Letter it is argued that the linear-zigzag transition is a quantum phase transition, which can be mapped
to the one of an Ising chain in a transverse field, describing a
quantum ferromagnetic transition at $T=0$~\cite{QPTbook}. This
equivalence was previously proposed in a theoretical study of
linear-zigzag structures for electrons in quantum
wires~\cite{MML}. Here we outline a mapping which allows us to
estimate the quantum critical point in terms of the system
parameters, and determine the experimental conditions required to
observe it. The
phase diagram is sketched in Fig.~\ref{Fig:1}.
We find a finite, measurable deviation from the critical point
predicted by the classical theory~\cite{FCCM}.

The system we consider is composed of $N$ ions of mass $m$ and
charge $Q$. The ions mutually repel via the Coulomb
interaction and are confined in the transverse plane ($y-z$) by an
anisotropic harmonic potential, while periodic boundary conditions
are assumed in the $x$ direction. The ions experience the
potential  $V_{\rm tot}=V_t+V_{\rm int}$, which is the sum of the
transverse harmonic trap, $V_t=m\nu_t^2\sum_{j=1}^N y_j^2/2$, and
the Coulomb repulsion, $V_{\rm
int}=(Q^2/2)\sum_{i\not=j}1/\sqrt{(x_i-x_j)^2+(y_i-y_j)^2}$, while
the motion in the $z$ direction is frozen out by a tight confining
potential, so that the system is essentially two dimensional. The
parameters of the potential are
such that the particles are aligned along the $x$-axis with uniform interparticle distance
$a$, and their equilibrium positions are $x_j^{(0)}=ja$,
$y_j^{(0)}=0$, with $j=1,\ldots,N$. This configuration can be
realized in a ring trap
under the condition that one ion is pinned~\cite{Walther}, or in a anharmonic
trap~\cite{Monroe}. It also 
approximates the ions distribution at the center of a long chain in a linear
Paul trap~\cite{Walther,Drewsen,FCCM,MF}.

At ultralow temperatures the ions perform harmonic vibrations
about the mechanically stable equilibrium positions. The linear
chain is stable against longitudinal displacements from the
equilibrium positions $x_j^{(0)}$, while a soft mode in the
transverse direction drives a structural transition into a zigzag
configuration. This mode has wavelength $\lambda=2a$, such that in
the classical harmonic crystal the corresponding displacements are
$y_j^{\rm soft}=(-1)^jy_0$ with $y_0$ the amplitude of the
oscillations.  We focus on the dynamics of the transverse phonon
modes of the linear chain close to the instability, and expand
$V_{\rm int}$ to fourth order in $|y_j|\ll a$. Our starting point
is the Landau model derived in Ref.~\cite{FCCM}. Close to the
instability the displacement along $y$ can be rewritten as
$y_j=(-1)^j a\phi_j$, where $\phi_j$ is a dimensionless field
which varies slowly with the ion position. To leading order in a
gradient expansion, we obtain an effective potential for the field
$\phi_j$
\begin{equation}
V[\{\phi_j\}]\approx \sum_{j=1}^{N} V_0(\phi_j) + \frac{1}{2}K\sum_{j=1}^N (\phi_j-\phi_{j+1})^2\; ,
\label{V_phi}
\end{equation}
where $V_0$ is a local potential of the form
\begin{equation}
V_0(\phi)=-\frac{1}{2}m(\nu_c^2-\nu_t^2)a^2\phi^2+\frac{1}{4}ga^4\phi^4\; ,
\label{V_loc}
\end{equation}
and the second term in Eq.~(\ref{V_phi}) describes
nearest-neighbor interaction at strength $K=(Q^2/a)\ln 2$. The
other parameters are $\nu_c=\sqrt{4Q^2/ma^3}\sqrt{C_3}$ and $g=24
C_5 Q^2/a^5$, with $C_\alpha\equiv \sum_{j\geq 1}
1/(2j+1)^{\alpha}$. Note that $\nu_c$ is the frequency at which
the linear chain is unstable and a transition to the zigzag
configuration occurs within the classical
theory~\cite{Dubin93,FCCM}. It is remarkable that the Coulomb
repulsion is mapped to a short-range potential close to the
linear-zigzag instability. This property is clearly visible in the
spectrum of the
transverse excitations: $\omega_\perp^2(k) \sim
\nu_t^2-\nu_c^2+(K/m)k^2$~\cite{FCCM}, characteristic of a chain
with nearest-neighbor coupling and sound velocity $c=\sqrt{K/m}$.

We next define a parameter
$\varepsilon\equiv(\nu_c^2-\nu_t^2)/\nu_c^2$. Classically, the
chain is stable for $\varepsilon<0$ \cite{FCCM}. For $\varepsilon
>0$, the local potential $V_0$ in Eq.~(\ref{V_loc}) has minima at
$\phi_\pm=\pm \phi_0$, with
$\phi_0\equiv(\nu_c/a)\sqrt{m\varepsilon/g}$; the ions therefore
order in a zigzag structure at either one of the configurations
with $y_j^{(\pm)}=(-1)^j\phi_\pm a$, thereby breaking the
reflection symmetry at the $x$-axis. This calculation, however,
neglects quantum fluctuations, which induce tunnelling between the
minima and tend to destroy the zigzag order.

Quantum fluctuations are included in the model of the structural transition by means of a 1+1 dimensional quantum field theory \cite{long}. To this end, we write the partition function as
\begin{equation}
Z=\int {\mathcal D}\phi\, e^{-S[\phi]/\hbar}
\label{Z_def}
\end{equation}
with the Euclidean action
\begin{eqnarray}
S[\phi]=\int_0^\beta d\tau \sum_{j=1}^N
\left[\frac{ma^2}{2}(\partial_\tau \phi_j)^2
+V_0(\phi_j)+\frac{K}{2}(\phi_j-\phi_{j+1})^2\right]
\label{S_phonons}
\end{eqnarray}
where $\beta=\hbar/k_BT\rightarrow \infty$ for $T\rightarrow 0$.
This model describes the quantum dynamics of the zigzag phonon mode in terms of a real continuous scalar field, $-\infty<\phi_j(\tau)<\infty$.
We now express it in a form which allows its mapping onto an effective model for the {\it discrete} field $\sigma_j(\tau)\equiv {\rm Sgn}[\phi_j(\tau)]$.
We first divide the imaginary time ($\tau$) axis into $M$ discrete steps separated by an infinitesimal interval of size
$\delta\tau=\beta/M$. The partition function is then cast in the form $Z={\rm Tr}\,\{\hat T^M\}$, where $\hat T$ is a transfer matrix describing the time evolution from $\tau$ to  $\tau^\prime=\tau+\delta\tau$, whose elements read
\begin{eqnarray}
& &T[\{\phi_j\},\{\phi^\prime_j\}]=\prod_{j=1}^NT_{loc}(\phi_j,\phi^\prime_j;\delta\tau)\label{T_mat}\\
&\times &\exp\left\{-\frac{\delta\tau}{4\hbar}K
\sum_{j=1}^N\left[(\phi_j-\phi_{j+1})^2+(\phi^\prime_j-\phi^\prime_{j+1})^2\right]\right\}.
\nonumber
\end{eqnarray}
Here, $T_{loc}(\phi_j,\phi^\prime_j;\delta\tau)=\langle
\phi^\prime_j |e^{-{\hat H_j}\delta\tau/\hbar}|\phi_j\rangle$ is
the propagator describing the quantum dynamics of a particle
subject to the local Hamiltonian $\hat
H_j=p_j^2/(2m)+V_0(\phi_j)$, with $V_0(\phi)$ the double--well
potential, Eq.~(\ref{V_loc}), and $\phi_j,p_j$ canonical conjugate
variables.
We now write the continuous field as
$\phi_j=\phi_0\sigma_j^z+\delta\phi_j$, where $\sigma_j^z$ is the
Pauli matrix, which can take values $\pm 1$ corresponding to the
minima of the double-well potential, and $\delta\phi_j$ are the
fluctuations. We then integrate over $\delta\phi_j$, obtaining an
effective transfer matrix in terms of Ising fields. When the
system is well inside the zigzag phase, the typical frequency of
tunnel-splitting in the double-wells, $\Delta\omega$, is much
smaller than $\omega_0$, the frequency of the oscillations about
the well minima. Under this approximation, we obtain
\begin{eqnarray}
& & T_{eff}[\{\sigma_j\},\{\sigma^\prime_j\}]
\label{Teff}\\&=&\prod_{j=1}^N T_{loc}[\sigma_j,\sigma^\prime_j]
\exp\left\{\frac{K\phi_0^2\delta\tau}{2\hbar}\left(\sigma_j\sigma_{j+1}+\sigma^{\prime}_j\sigma^{\prime}_{j+1}\right)
\right\}\nonumber
\end{eqnarray}
where the $\,2\times 2$--matrix ${\hat T}_{loc}$ encodes the single particle dynamics within the double-well potential. The propagator ${\hat T}_{loc}$ obeys the symmetry $T_{loc}[\sigma,\sigma^\prime]=T_{loc}[-\sigma,-\sigma^\prime]$, and can therefore be written as
\begin{equation}
{\hat T}_{loc}=const\times (A\sigma^0+B\sigma^x)\; ,
\label{Tloc_gen}
\end{equation}
where $\sigma^\alpha$ are Pauli matrices in the basis where
$\{\sigma=\pm 1\}$ denote the eigenvalues of $\sigma^z$, and
$\sigma^0$ is the $\,2\times 2$ unit matrix. The coefficients $A$
and $B$ acquire a simple form in the zigzag phase, when the ion
wave functions are relatively well localized near $\pm\phi_0$ and
can be approximated by Gaussians of width $l$. In this limit, using the
inequality $e^{-2\phi_0^2/l^2} \ll\tanh\{\Delta\omega\delta\tau
/2\}\ll 1$, we obtain
\begin{eqnarray}
{\hat T}_{loc}\approx
T_0\exp\left\{\Delta\omega\sigma^x\delta\tau/2\right\}
\label{Tloc_sigma_x}
\end{eqnarray}
with $T_0$ a constant prefactor. As a result, the partition function reduces to the form $Z\approx Z_0\,\int {\mathcal D}\sigma {\rm exp}\left(-S_I[\sigma]/\hbar\right)$, in which $Z_0$ is a constant and $S_I$ the action of an Ising chain in a transverse field \cite{QPTbook} subject to the Hamiltonian
\begin{equation}
H_I=-\sum_{j=1}^N (J\sigma^z_j\sigma^z_{j+1}+h\sigma^x_j)\; .
\label{H_Ising}
\end{equation}
Here, the parameters are the fictitious exchange coupling
\begin{equation}
J=K\phi_0^2=C_JU_P\varepsilon,
\end{equation}
with $U_P=Q^2/a$ the strength of the interaction and $C_J=C_3\ln 2/(6C_5)$, while the transverse field $h=\hbar\Delta\omega/2$ is proportional to the splitting energy. Its dependence on the physical parameters can be estimated by means of a variational calculation \cite{long}, which gives
\begin{equation}
h\approx C_h\left(U_PU_K^2\right)^{1/3}
\label{Ising_Jh}
\end{equation}
where $U_K=\hbar^2/ma^2$ is the typical kinetic energy scale of the atoms in the chain and $C_h=(9C_5/2)^{1/3}$. Equation~(\ref{Ising_Jh}) was obtained assuming $U_K/U_P\gg \varepsilon^3$, which holds close to the critical points, where quantum fluctuations dominate.

The model described by Hamiltonian~(\ref{H_Ising}) is known to exhibit a quantum phase transition at $h/J=1$ and $T=0$, separating
an ordered phase at $J>h$ (in our case, the stable zigzag configuration) from a disordered phase at $J<h$ (the linear chain). In both phases the spectrum of excitations is characterized by a gap with energy $\Delta=2|J-h|$. The corresponding phase diagram,
showing the critical behavior as a function of $\varepsilon$ (and hence $\nu_t$) and $T$ is sketched in Fig.~\ref{Fig:1}. The critical regime
($k_B T>\Delta$) is characterized by universal power-law $T$-dependence of correlation functions with the critical exponents of the quantum transition~\cite{QPTbook}.

The
mapping to the Ising model, here presented, is strictly
valid as long as the lowest energy levels of the local,
double-well potential are below the barrier. In this regime, the
tunneling between the wells is small and hence necessarily $h\ll
J$, which is satisfied deep in the zigzag phase. The symmetry of
the ordered phase was used to identify the effective model and
estimate its parameters. This procedure is strictly valid only
in a higher dimension ($4-\epsilon$).  Nevertheless, from conformal
symmetry arguments~\cite{Cardy} it is reasonable to assume that a quantum Ising model is the appropriate field theory for the $1+1$-D $\phi^4$-model describing the present system, and qualitatively describes its critical behavior at $h\sim J$. This hypothesis is further supported by numerical studies~\cite{Fisher}. Under this assumption we estimate the quantum critical point in
the regime of parameters where classical zigzag order is suppressed by quantum tunneling, and observe that it belongs to a universality class~\cite{QPTbook,Cardy} that differs from the one of the classical Landau model.

The critical value of $\varepsilon$, and thus of the transverse frequency $\nu_t$, is dictated by the condition $h=J$, yielding
\begin{equation}
\varepsilon_c\approx C_c\left(U_K/U_P\right)^{2/3} ,\; C_c\equiv C_h/C_J\sim 10\; .
\label{qcp}
\end{equation}
Hence, for given $U_K/U_P\ll 1$, a quantum critical point is
expected at $\varepsilon_c\gg U_K/U_P$.

We now discuss what are the experimental conditions which are
required in order to observe the quantum critical point. Chains
with dozens of ions are usually realized in linear Paul
traps~\cite{Walther,Wineland,Drewsen}, where the ion distribution
along the chain is inhomogeneous. The considerations here reported
apply in the center of the chain, when the chain contains several
dozens of ions.  More regular distributions are expected in linear
traps where the axial potential is anharmonic~\cite{Monroe}. Here,
one can tune through the critical point by either controlling the
transverse frequency $\nu_t$ (typically in the MHz regime), or the
spacing between neighboring ions $a$ (typically several $\mu$m) by
means of the axial confinement. To be able to distinguish the
quantum disordered phase from the ordered (zigzag) phase, the
frequency difference $\delta\nu=\nu_c-\nu_t$ should exceed the
experimentally accessible resolution. In order to estimate the
required resolution, we write $\varepsilon_c\approx 2 \delta\nu/\nu_c$ in terms of
$a=a_0\times 1\,\mu$m and of the ion mass $m=n_Am_p$, with $m_p$ the proton mass and $n_A$ the atomic
number, and find $ \varepsilon_c\approx 10^{-4}/(n_Aa_0)^{2/3}$,
leading to an upper bound on the frequency resolution
\begin{equation}
\delta\nu\approx 10^{-4}(\nu_c/2)(n_Aa_0)^{-2/3}
\label{delta_nu}
\end{equation}
ranging from KHz (for protons with $a_0\sim 1$) to several $\,$Hz for, e.g., $^{24}$Mg$^+$ ions. This bound must also be compared with the heating time scale $T_{\rm h}$ in ion traps, such that  $\delta\nu T_{\rm h}\ll 1$ should be satisfied, which leads to demanding conditions for the existing trapping setups~\cite{Ibk_Review}. A larger value $\delta\nu$, and hence less restrictive conditions on observing the quantum critical point, could be reached in presence of screening, when the chain is embedded in a crystal of another ion species~\cite{Drewsen}. It is also essential to reduce  $T$ below the energy scale characteristic of the gap away from criticality, i.e., $\Delta\sim h$, resulting in $T[{\rm mK}]\ll 10^{3}C_h(U_K^2U_P)^{1/3}/k_B$, which corresponds to
\begin{equation}
T[{\rm mK}]\ll 0.25\left(n_A^2a_0^5\right)^{-1/3}\,,
\label{T_num}
\end{equation}
implying an upper bound of order $\sim 0.1$mK for protons to several $\mu$K for Mg$^+$ ions.

The quantum critical point could be more easily observed in a quasi one-dimensional array of ultracold polar molecules, interacting via dipole-dipole repulsion, when the dipoles are aligned by an external field in a direction perpendicular to the plane where their motion takes place~\cite{Lahaye,Kollath}. Numerical simulations show that quantum tunneling clearly modifies the behavior at the linear-zigzag transition predicted by the classical theory~\cite{Astrakharchik}. Assuming the dipoles are pinned by an optical lattice at fixed interparticle position, then the potential for the phononic modes at the instability is given by Eq.~(\ref{V_phi}), and the mapping to the Ising model in the transverse field can be performed, where now $U_P=C_{dd}/(4\pi a^3)$, with $C_{dd}$ the dipole interaction strength~\cite{Astrakharchik}, while $C_c\sim 10$. Using typical numbers for the dipolar strength~\cite{Lahaye} one finds $\delta\nu\sim \nu_c$, showing that quantum fluctuations are  significant at the linear-zigzag transition.

The phase transition can be experimentally measured by light
scattering
via the structure factor $S(k)$
~\cite{Astrakharchik}. In the critical region
$S(k)=S_f(k)+S_0\delta(k-k_0)$, where $k_0=\pi/a$ is the wave
number of the zigzag, $S_0$ is proportional to the squared order
parameter, $S_0\propto \phi_0^2$, while $S_f$ is proportional to
the isothermal susceptibility $\chi$ of the corresponding Ising
system~\cite{QPTbook}. Approaching the quantum critical point in
the zigzag phase ($\epsilon> \epsilon_c$) and for $T\ll\Delta$,
one has $\chi\sim |\epsilon_c-\epsilon|^{-\gamma}$ and
$\phi_0^2\sim |\epsilon_c-\epsilon|^{2\beta}$ where $\beta=1/8$
and $\gamma=7/4$ are the exponents of the classical two
dimensional Ising model. In the critical region $T\gg \Delta$
(shaded area in Fig. 1), $\chi\sim T^{-7/4}$ and $\phi_0^2\sim
T^{1/4}$ ~\cite{QPTbook}. Note that this critical behavior is
strictly valid for an infinite system and assumes a uniform
density of ions. It is expected to be approximately valid in an
ion chain, when the variation of the density in a large region is
negligible within a correlation length.

To conclude,
we argue that the linear-zigzag instability in two-dimensional systems of trapped ions or polar molecules can be mapped to the one dimensional Ising model in a transverse field. This result demonstrates once more the potentialities offered by these systems as quantum simulators~\cite{Ibk_Review,Lahaye,QMagnet}, and more generally for quantum technological applications.

We acknowledge discussions with E. Altman, E. Demler, J. Eschner,
R. Fazio, J. Feinberg, Y. Gefen, J. Meyer, M.
Raizen, and S. Sachdev. This work has been partially supported by
the
ISF, the
BSF, the Minerva Center of Nonlinear Physics of Complex Systems,
MOST Grant No. 3-5792, the German Research Foundation (DFG), the
European Commission (FP7 2007-2013, STREP PICC, AQUTE), and the
spanish Ministery of Science and Innovation (FIS2007-66944;
EUROQUAM ``CMMC``).

\end{document}